\documentclass[aps,prfluids,reprint,onecolumn]{revtex4-2} 

\usepackage{graphicx}
\usepackage{dcolumn}
\usepackage{bm}

\usepackage{xcolor}
\usepackage{amsmath}
\usepackage{siunitx}

\newcommand{\pd}[2][]{\partial_{#2} #1}

\newcommand{\pddd}[2][]{\partial_{#2}^3 #1}

\newcommand{\integ}[4]{\int_{#1}^{#2}{\, #3 \, \mathrm{d} #4}}

\def\*#1{\bm{#1}}
\newcommand{\e}{\mathrm{e}}

\newcommand{\In}{\text{in}}
\newcommand{\Out}{\text{out}}

\begin{document}
\title{Droplet motion on chemically heterogeneous substrates with mass transfer. I. Two-dimensional dynamics }
\author{Danny Groves}
\affiliation{School of Mathematics, Cardiff University, Cardiff, CF24 4AG, UK}
\author{Nikos Savva}
\email{Corresponding author: n.savva@cyi.ac.cy}
\affiliation{Computation-based Science and Technology Research Center, The Cyprus Institute, Nicosia 2121, Cyprus}

\begin{abstract}
We consider the dynamics of thin two-dimensional viscous droplets on chemically heterogeneous surfaces moving under the combined effects of slip, mass transfer and capillarity. The resulting long-wave evolution equation for the droplet thickness is treated analytically via the method of matched asymptotic expansions in the limit of slow mass transfer rates, quasistatic dynamics and vanishingly small slip lengths to deduce a lower-dimensional system of integrodifferential equations for the two moving fronts. We demonstrate that the predictions of the deduced system agree excellently with simulations of the full model for a number of representative cases within the domain of applicability of the analysis. Specifically, we focus on situations where the mass of the drop changes periodically, to highlight a number of interesting features of the dynamics, which include stick-slip, hysteresis-like effects, as well as the possibility for the droplet to alternate between the constant-radius and constant-angle stages which have been previously reported in related works. These features of the dynamics are further scrutinized by investigating how the bifurcation structure of droplet equilibria evolves as the mass of the droplet varies.
\end{abstract}

\maketitle

\section{Introduction}\label{Sec:Introduction}

Moving droplets on surfaces is an ubiquitous phenomenon in the natural world, which also finds application in a broad spectrum of technologies, such as in hydrogen fuel cells \cite{Kim2014,Cheah2013}, DNA analysis \cite{Dugas2005}, printing  \cite{Park2006,Calvert2001}, as well as in the fabrication of display technologies \cite{Eales2015}, to name a few. Yet, their study is inherently complex and a lot of these applications are being developed based on intuition derived from laboratory observations. This complexity stems from the multi-scale nature of the phenomenon, from forces that manifest themselves at the macroscale, such as gravity and capillarity, to the microscale effects close to the droplet front \cite{Snoeijer2013}. Although impressive progress in contact line phenomena and the broader field of wetting hydrodynamics has been made in recent decades \cite{Bonn2009}, there are several open questions that still remain, including the poorly understood contact line dynamics on surfaces decorated with chemical and/or topographical heterogeneities. When these are present, many interesting phenomena emerge, including the pinning of the fronts \cite{Cubaud2004,Bonn2009}, stick-slip effects \cite{Rio2006,Chung2007,Kusumaatmaja2007}, and contact angle hysteresis  \cite{Eral2012,Brandon1996,Kusumaatmaja2007}. In most circumstances these heterogeneities occur naturally and are unavoidable, but there are also situations in nature and technology where surface texturing is desirable, e.g., to facilitate the self-cleaning of plant leaves and artificial surfaces \cite{Barthlott1997,Parkin2005}, to enhance condensation \cite{Ghosh2014,Lee2012} or for directed droplet transport in microfluidic and lab-on-a-chip devices \cite{Pravinraj2019}.

Thanks to sophisticated methodologies which have emerged in recent years, it is now possible to control these features at minute lengthscales (see, e.g., Refs.~\cite{Zhou2018,Baek2018}). For example, when a surface is decorated with alternating wettability patterns \cite{Bliznyuk2009,Jansen2012,Jansen2014,David2012} droplets tend to align themselves to the features of the surface, preferentially spreading along the more wettable regions; for sufficiently stronger wettability contrasts it is also possible for droplets to split into satellite bodies \cite{Zou2018}. For this reason, there is strong interest in improving our fairly rudimentary understanding of how surface heterogeneities impact droplet motion, by developing predictive models that allow us to more precisely control droplet behavior in the aforementioned applications and beyond.

Interesting dynamics also manifest themselves when a droplet is subjected to change in its mass which may occur due to a variety of mechanisms, including, e.g., pumping liquid into the droplet \cite{Lam2002}, liquid imbibition through a permeable substrate \cite{Arora2006,Aydemir2010,Espin2015},  evaporation (see, e.g., Refs. \cite{Brutin2018,Erbil2012} for reviews), or a combination thereof together with other effects (see, e.g., Refs.~\cite{Pham2019,Charitatos2021}). During mass transfer, two distinct modes or a combination thereof are reported primarily in the context of evaporative dynamics \cite{Erbil2012}, namely the \emph{constant-radius mode}, during which the contact line appears to be pinned, and the \emph{constant-angle mode}, i.e. when the radius evolves such that the apparent contact angle remains constant \cite{brutin2015droplet,Stauber2015,Stauber2015a,Amini2017, Amini2017a,Pham2017,Erbil2002,Armstrong2019}. Noteworthy also is the recently reported \emph{snapping mode}, whereby a droplet transitions to different locations on textured surfaces with macroscopic features while losing mass due to evaporation, but on a slower timescale compared to stick-slip dynamics~\cite{Wells2018}.

In the present study we aim to investigate the coupled macro- and microscale problem of a droplet spreading over a chemically heterogeneous surface, which is additionally subjected to changes in its mass. Mass transfer effects are prescribed by an arbitrary spatiotemporal function and may equivalently model imbibition through the substrate, or mass transfer through the free surface of the droplet \cite{Kiradjiev2018,OLIVER2015}. This investigation is performed by considering the corresponding long-wave evolution equation for the droplet thickness with a slip condition, which can be derived by standard arguments in the limit of small slopes, strong surface tension and negligible inertia \cite{Greenspan1978,HOCKING1983,OLIVER2015,Kiradjiev2018}. Analytical progress is possible in the vanishingly small slip limit by assuming that the spreading of the droplet and mass changes are sufficiently slow so that the method of matched asymptotic expansions can be leveraged to uncover the coupling of the dynamics at the macroscale where capillary forces dominate, and the microscale close to the contact line where slip effects feature more prominently \cite{Lacey1982,HOCKING1983}.

The discussion here is limited to the two-dimensional (2D) geometry  primarily to highlight, in qualitative terms, the interesting interplay between surface heterogeneities and mass transfer, deferring the arguably more realistic three-dimensional (3D) setting to the following paper \cite{Savva2021}. We also seek to extend related works in 2D on homogeneous surfaces, e.g. the work of Oliver \emph{et al.} which focused on the particular case of a constant mass flux, but, unlike here, a variety of distinguished limits were considered  \cite{OLIVER2015}, or that of \citeauthor{Kiradjiev2018} which looked into symmetric motion when the mass flux is localised at the center of the drop \cite{Kiradjiev2018}. Apart from the inclusion of chemical heterogeneities, two additional key features of the present theory is that it is formulated for mass fluxes that are arbitrary functions of the independent variables, and that it includes higher-order correction terms in the asymptotics of the distinguished limit considered here. Noteworthy also are related numerical studies of the long-wave modeling of imbibition dynamics on permeable substrates \cite{Espin2015} and on localized fluxes on heterogeneous surfaces within the framework of the computationally more demanding diffuse interface Cahn--Hilliard formalism \cite{Pradas2016}. Mass transfer effects can also arise in the context of evaporative dynamics, but, unlike the present study, mass fluxes are maximized in the vicinity of the contact line. These include work in the diffusion-limited regime, which gives rise to mass fluxes that are weakly singular near the contact line \cite{Saxton2016,Saxton2017}, as well as evaporation in a pure-vapor atmosphere on homogeneous \cite{Savva2017,Amini2017} and rough surfaces \cite{Amini2017a,Pham2017,Charitatos2021}, where the mass flux can be expressed as a function of the droplet thickness.

The presentation of the work is organized as follows. In Sec.\ \ref{Sec:ProblemFormulation} we briefly describe the assumptions and governing equation for our model. In Sec.\ \ref{Sec:MA} we present the matched asymptotic analysis undertaken by assuming that motion occurs quasistatically to determine a set of evolution equations for the contact line velocities which arise through the asymptotic matching. In Sec.\ \ref{Sec:Results} we explore the predictions found through the reduced models and assess how well they compare with simulations of the full governing partial differential equation (PDE). In Sec.\ \ref{sec:bifurcations} we further scrutinize the effect of mass flux on droplet motion by invoking a bifurcation analysis with numerical continuation techniques to explore apparent hysteresis and other effects. A number of concluding remarks are offered in Sec.\ \ref{Sec:Conclusion}.

\section{Problem formulation}\label{Sec:ProblemFormulation}

\begin{figure}
	\includegraphics[scale = 0.9]{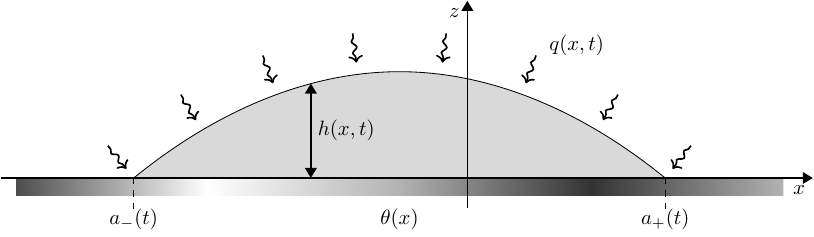}
	\caption{The problem geometry in dimensionless units. The droplet moves on a chemically heterogeneous substrate with spatially varying contact angles given by $\theta(x)$. The droplet profile is described by its thickness $h(x,t)$ and its right/left contact points, given by $a_\pm(t)$. The droplet is subjected to changes in its mass, as prescribed by mass flux $q(x,t)$, which may occur either through the free surface or through imbibition along the substrate.}\label{fig:geometry}
\end{figure}

Consider a 2D droplet moving over a flat, horizontal and chemically heterogeneous substrate. The droplet profile is described at position $x$ and time $t$ through its thickness $h(x,t)$ (see Fig.~\ref{fig:geometry}). Assuming small slopes everywhere, strong surface tension effects, negligible inertia and gravity, a governing PDE is deduced for the evolution of the droplet thickness by performing a small slopes expansion of the Stokes equations \cite{OLIVER2015}, namely 
\begin{subequations}\label{full}
\begin{equation}\label{GovPDE}
	\pd[h]{t}+\pd[]{x}\left[h(h^2+\lambda^2)\pddd[h]{x}\right]=q(x,t).
\end{equation}
Here, Eq.~\eqref{GovPDE} is cast in dimensionless form, where $\lambda$ is the slip length, and $q(x,t)$ is the flux which may be attributed to mass transfer occurring at the free surface of the drop via the kinematic boundary condition \cite{OLIVER2015}, or through the substrate \cite{Kiradjiev2018}.  Slip is modelled according to the inverse linear slip boundary condition as proposed by Ruckenstein and Dunn \cite{Ruckenstein1977} (see also Ref.~\cite{Greenspan1978}) which removes the stress singularity associated with moving contact lines~\cite{Moffatt1964,Huh1971}. Let us note that we have opted against the arguably more popular Navier slip model because the pressure is logarithmically singular along the contact line, therefore complicating its numerical implementation \cite{Savva2009}, particularly in the 3D counterpart of the present study \cite{Savva2021}. The droplet touches the substrate at $x=a_\pm(t)$ which are, respectively, its right and left contact points (see Fig.~\ref{fig:geometry}), expressed through the conditions
\begin{equation}\label{Cond1}
	h(a_\pm,t) = 0.
\end{equation}
Additionally, the droplet meets the substrate at the local contact angle, namely by requiring
\begin{equation}\label{Cond2}
	\left.\pd[h]{x}\right|_{x=a_{\pm}} =\mp\theta(a_\pm) = \mp\theta_{\pm},
\end{equation}
where $\theta(x)$ is a spatially varying function that prescribes the local contact angles due to the chemical heterogeneities, assumed to be small. There are no additional assumptions on $\theta(x)$, only that it varies at length-scales much longer than $\lambda$. Since this is a free-boundary problem, it is also necessary to derive a set of conditions that are defined compatibly with
\begin{equation}\label{AreaConstraint}
	\frac{\mathrm{d}}{\mathrm{d} t}\integ{a_-}{a_+}{h}{x} = \integ{a_-}{a_+}{q}{x} = \dot{v}(t),
\end{equation}
found by integrating directly \eqref{GovPDE} from $a_-$ to $a_+$, where dots denote differentiation with respect to time and $v(t)$ is the cross sectional area of the droplet (denoted by $v$ to be consistent with the notation for the 3D volume in Part II of the present investigation~\cite{Savva2021}). Through a local expansion of \eqref{GovPDE} and making use of \eqref{Cond1} and \eqref{Cond2} (see, e.g.\ Ref.~\cite{OLIVER2015}), we deduce the following free-boundary conditions
\begin{equation}\label{clbcs}
	\dot{a}_\pm = \lim_{x\to a_\pm} \left(\lambda^2 \partial_x^3h\pm \frac{q}{\theta}\right).
\end{equation}
\end{subequations}
In this manner, the problem specified by \eqref{full} determines the dynamics of $h$ and $a_\pm$ in the wetted region $a_-\leq x \leq a_+$ subject to an appropriately chosen initial condition. The Eqs.\ \eqref{full} are non-dimensional according to the scalings
\begin{equation}
	X = xL,\quad T = \frac{3\mu Lt}{\sigma\alpha_s^3},\quad A_\pm=La_\pm,\quad H = hL\alpha_s,\quad \Lambda = \frac{\lambda L\alpha_s}{3},\quad Q = \frac{q\sigma \rho \alpha_s^4}{3\mu},\quad \Theta = \alpha_s \theta, \label{units}
\end{equation}
with the lower-case letters denoting the dimensionless counterparts of the capital ones. Here $\rho$ is the fluid density, $\sigma$ the surface tension, $\mu$ the fluid viscosity, and $\alpha_s$ corresponds to a (generally small) reference contact angle, e.g.\ taken to be some characteristic contact angle; in the case of substrates with localized features $\alpha_s$ corresponds to the contact angle in the defect-free region of the substrate. Finally, the lengthscale  $L = \sqrt{V/(2\alpha_s)}$ is a measure of the characteristic size of the droplet and is associated with its cross-sectional area.

Solving \eqref{full} numerically as $\lambda\to 0$ (since $\lambda\ll1$ for macroscopically large droplets) causes numerical stiffness to increase because of the difficult-to-resolve boundary layers for $\pd[h]{x}$ near the contact line, as we have a transition from the apparent contact angle in the bulk to the prescribed angles at the contact line. Therefore, to provide an attractive alternative to numerical calculations, as well as to afford deeper physical insights to such a complex multi-scale problem, we invoke the use of matched asymptotics as in related works~\cite{HOCKING1983,Savva2009,Vellingiri2011,OLIVER2015,Savva2017}.

\section{Matched asymptotic analysis} \label{Sec:MA}

The analysis is undertaken in the limit $\lambda\to0$. It closely follows previous works that treated the case when $q\equiv0$, whereby the domain of interest is split into an outer region that makes up the bulk of the droplet and two inner regions in the vicinity of $x=a_\pm$ \cite{Lacey1982,HOCKING1983,Vellingiri2011}. Within the outer region, the dynamics is dominated by capillary and viscous forces, with slip effects becoming more predominant near the contact points. The aim is to obtain a set of evolution equations for the contact point velocities, $\dot{a}_\pm$, which is accomplished by linking the macro- with the microscale dynamics by means of asymptotically matching the respective solutions in each of these regions.

This approach becomes feasible by assuming that there is a separation of scales (e.g., for droplets sufficiently far from equilibrium and whose size is much larger than $\lambda$) and treating the dynamics as being quasisteady. This is a realistic assumption for many situations of practical interest and corresponds to the surface-tension dominated regime of small capillary numbers, i.e.\ $Ca=\mu U/\sigma\ll1$, with $U$ being a characteristic velocity scale \cite{Bonn2009}. This ultimately allows us to delay the contributions of the $\dot{a}_\pm$ terms to appear in the next-to-leading term, ultimately giving rise to the scalings $|\dot{a}_\pm| = O\left(1/|\ln \lambda|\right)\ll1$ as $\lambda\to0$ \cite{Lacey1982,HOCKING1983}. The present analysis maintains these scalings for $\dot{a}_\pm$, by focusing on the distinguished limit for which $q$ and, consequently, $\dot{v}$ are also $O(1/|\ln\lambda|)$ as $\lambda\to0$. In this manner, the analysis  remains more tractable compared to other distinguished limits, without compromising any of the qualitative features of the dynamics we wish to uncover. Should these flux terms be present at leading order, a separate treatment would be required (see Ref.~\cite{OLIVER2015} for other distinguished limits when $\theta(x)\equiv 1$ and $q$ is constant). 

As $\lambda \to 0$ slip effects are negligible in the outer region and thus the slip parameter, $\lambda$, is dropped from the leading-order dynamics of Eq.~\eqref{GovPDE}. To simplify the analysis that follows, we map the time dependent wetted domain $a_-(t)\leq x \leq a_+(t)$ of Eq.~\eqref{full} to a fixed one $-1\leq s \leq 1$ using the change of the variable
\begin{equation} \label{CoordMap}
	x = \frac{1}{2}\big[(a_+-a_-)s+a_++a_-\big],
\end{equation}
so that the PDE for the outer solution, $h_\Out$, is transformed in the limit $\lambda \to 0$ to
\begin{equation}
	\pd[h_\Out]{t}-\frac{\dot{a}_{+}(1+s)+\dot{a}_-(1-s)}{2d}\pd[h_\Out]{s}+\frac{1}{d^4}\pd[]{s}\left(h_\Out^3\pddd[h_\Out]{s}\right) = q(s,t),
\end{equation}
where we define $d=(a_+-a_-)/2$ as the droplet half-width. The pertinent conditions for the outer region are
\begin{equation}
	h_\Out(\pm 1,t) =0,\quad\text{and}\quad \integ{-1}{1}{h_\Out}{s} = \frac{v}{d}.
\end{equation}
Just as in many aforementioned works, the dynamics occurs sufficiently slowly so that a quasistatic approximation is assumed to hold. This allows us to drop the explicit time dependence from $h_\Out(s,t)$ by introducing an expansion of the form
\begin{equation}\label{OuterExpansion}
	h_\Out = h_0(s,a_{\pm},v)+h_1(s,a_{\pm},v,\dot{a}_{\pm},\dot{v})+\ldots,
\end{equation}
where $\dot{v}$ and $\dot{a}_\pm$ are assumed to be small as $\lambda \to 0$ and $h_0 \gg h_1$. In essence, $h_0$ describes the quasisteady  state of the droplet, and $h_1$  is linear in the $\dot{a}_\pm$ and $\dot{v}$ terms.  This ordering of terms is formally justified by rescaling time according to the slow time scale which is  $O(|\ln\lambda|)$ as $\lambda\to0$ \cite{Lacey1982,OLIVER2015}. Using Eq.~\eqref{OuterExpansion},  we deduce the following problem for the $O(1)$ leading order term 
\begin{equation}
\pd[]{s}\left(h_{0}^3\pddd[h_0]{s}\right)= 0,\quad h_0(\pm 1,t) = 0 \quad\text{and}\quad \integ{-1}{1}{h_0}{s} = \frac{v}{d},\label{OuterLeadingEq}
\end{equation}
which is easily solved to yield a parabolic profile of the form
\begin{equation}\label{LeadingOrderHeight}
	h_0(s,t) = \frac{3v}{4d}\left(1-s^2\right).
\end{equation}
Here we note that in the leading-order problem \eqref{OuterLeadingEq} we have neglected $q$ terms in accordance with the assumption that $q$ and $\dot{v}$ are $O\left(1/|\ln\lambda|\right)$ as $\lambda\to 0$. Should $q$ terms be included at leading order, they would contribute to the shape of the droplet and a different approach would be required \cite{OLIVER2015}. Next, the equation for $h_1$ satisfies
\begin{equation}\label{eq:h1}
\pd[h_0]{t}-\frac{\dot{a}_{+}(1+s)+\dot{a}_-(1-s)}{2d}\pd[h_0]{s}+\frac{1}{d^4}\pd[]{s}\left(h_0^3\pddd[h_1]{s}\right) = q(s,t),
\end{equation}
which is solved subject to the conditions
\stepcounter{equation}\begin{equation}\label{h1conds}
	h_1(\pm 1,t) = \integ{-1}{1}{h_1}{s} = 0,\quad\text{and}\quad \integ{-1}{1}{q}{s} = \frac{\dot{v}}{d},\tag{10a,b}
\end{equation}
where, for notational simplicity, we write $q$ to be dependent directly on the $s$ variable rather than indirectly through $x$ as transformed according to Eq.~\eqref{CoordMap}. As in related works, we seek to find the behavior of the slopes as $s\to \pm 1$ \cite{Savva2009,Vellingiri2011}. From a local expansion of the governing equation for $h_1$, we find that both $\dot{a}_\pm$ and $q_\pm=q(\pm1,t)$ contribute a logarithmic singularity as $s\to1$. The corresponding two-term asymptotic expansion for $\pd[h_1]{s}$ is cast in the form
\begin{equation}\label{h1behaviour}
	\pd[h_1]{s}\sim  - \frac{d(\vartheta\dot{a}_{\pm}\mp q_\pm)}{\vartheta^3}\ln(1\mp s)-\beta_\pm\quad\text{as}\quad s\to \pm 1,
\end{equation}
where $\vartheta$ is the apparent contact angle as computed from the leading-order shape, Eq.~\eqref{LeadingOrderHeight}, namely
\begin{equation}\label{ApparentContactAngle}
	\vartheta = \mp\frac{1}{d}\left.\pd[h_0]{s}\right|_{s=\pm 1} = \frac{3v}{2d^2},
\end{equation}
and $\beta_\pm$ are time-dependent functions which are determined in Appendix~\ref{BetaCalc} from the asymptotic analysis of Eq.~\eqref{eq:h1}, namely
\begin{equation}\label{beta1}
	\beta_{\pm} = \frac{d}{\vartheta^2}\left[\dot{a}_{\pm}\ln\frac{\e^2}{2}-\dot{a}_\mp\right]+\frac{d}{\vartheta^3}\left[I_\pm \mp\frac{3\dot{v}}{2d}\pm q_\pm\ln 2\right],
\end{equation}
where
\begin{equation}\label{Iintegral}
	I_\pm = \integ{-1}{1}{\left[\frac{1}{2}\ln\left(\frac{1+s}{1-s}\right)q\pm\left(\frac{q - q_\pm}{1\mp s}\right)\right]}{s}.
\end{equation}
Hence, recasting Eq.~\eqref{h1behaviour} in terms of the $x$ variable, we find that, as $x\to a_\pm$,
\begin{equation}\label{OuterSlope}
	\mp\pd[h_\Out]{x}\sim \vartheta\pm\left(\frac{\dot{a}_{\pm}\vartheta\mp q_{\pm}}{\vartheta^3}\right)\ln\left[\frac{\mp\left(x-a_\pm\right)}{2d}\right]\pm\frac{2\dot{a}_{\pm}-\dot{a}_\mp}{\vartheta^2}-\frac{3\dot{v}}{2d\vartheta^3}\pm\frac{I_\pm}{\vartheta^3},
\end{equation}
which specifies how the slope of the outer region behaves as the contact points are approached.

The corresponding asymptotics of the inner region follows similar arguments as in the equivalent problem with constant mass \cite{Vellingiri2011}. Hence, by appropriately scaling Eq.~\eqref{GovPDE} close to the contact points and introducing a quasistatic expansion similar in appearance to \eqref{OuterExpansion}, we determine the two-term asymptotic expansions
\begin{equation}\label{InnerSlope}
\mp\pd[h_\In]{x}\sim\theta_{\pm}\pm\left(\frac{\dot{a}_{\pm}\theta_\pm\mp q_\pm}{\theta_{\pm}^3}\right)\ln\left[\e\theta_{\pm}\frac{\mp(x-a_{\pm})}{\lambda}\right]\quad\text{as}\quad \frac{x-a_{\pm}}{\lambda} \to \mp\infty,
\end{equation}
which describe the behavior as we move from the contact points to the bulk of the droplet (see Appendix~\ref{InnerRegionCalc} for more details).

As in most problems considering the asymptotics of contact lines, we find that the $x$-dependent logarithmic terms of the outer \eqref{OuterSlope} and inner \eqref{InnerSlope} solutions cannot directly match. In many instances, matching is possible by considering the cubes of the slopes, which is justified through the presence of intermediate regions between the respective inner and outer regions \cite{HOCKING1983,Savva2009,Vellingiri2011}. For the current problem, this approach fails to work due to the presence of $q_\pm$ in the singular terms of Eqs.~\eqref{OuterSlope} and \eqref{InnerSlope}. However, matching is still possible, albeit through a much more elaborate analysis, like the one introduced by Lacey for homogeneous substrates without mass transfer \cite{Lacey1982}, and its extension by Oliver \emph{et al.} to problems with constant $q$~\cite{OLIVER2015}. Alternatively, a recent problem-independent generalization offered by Sibley \emph{et al.}\ allows us to circumvent the effort required through an integral which gives the functional forms of the inner and outer slopes that directly match within their respective overlap regions~\cite{Sibley2015}. Thus, when $q_\pm\neq0$, matching can be performed in this manner, giving rise to transcendental equations for $\dot{a}_\pm$, which require a more involved numerical treatment (see, Eqs.~\eqref{transcendental} in Appendix~\ref{MatchingFull}). Since our principal aim is uncovering the qualitative features of the dynamics, we chose not to pursue this case here. Therefore, we have assumed that $q_\pm = 0$ (i.e.\ no mass transfer at $x=a_\pm$), so that matching through the cubes of the slopes becomes possible. Although having $q$ vanish at $x=a_\pm$ may be too restrictive in some cases, e.g.\ when we have mass loss through evaporation, as $q$ is maximized there \cite{Saxton2016,Savva2017,Saxton2017}, it is appropriate for cases where mass flux is localized somewhere within the droplet's footprint \cite{Pradas2016,Kiradjiev2018}. Thus, matching using the cubes of the slopes yields explicit expressions for $\dot{a}_\pm$, namely
\begin{equation}
	\dot{a}_{\pm} = \pm\frac{\kappa_{\pm}\ln\left(\dfrac{2d\theta_{\mp}}{\e\lambda}\right)	+\kappa_\mp}{\ln\left(\dfrac{2d\theta_+}{\e\lambda}\right)\ln\left(\dfrac{2d\theta_-}{\e\lambda}\right)-1},\label{IDESystem}
	\end{equation}
where
	\begin{equation}\label{kappa}
	\kappa_\pm = \frac{\vartheta^3-\theta_{\pm}^3}{3}\pm \frac{I_{\pm}}{\vartheta}-\frac{3\dot{v}}{2d\vartheta}.
	\end{equation}
The structure of this system of integrodifferential equations (IDEs) is the same as that obtained by Vellingiri \emph{et al.}\ in the absence of mass transfer \cite{Vellingiri2011}. The reduced system~\eqref{IDESystem} is arguably easier to solve compared to the full PDE, Eq.~\eqref{GovPDE}, especially in the limit $\lambda\to 0$ due to the numerical stiffness issues highlighted previously.

Without loss of generality, the flux is prescribed as $q=\dot{v}\tilde{q}$  so that Eq.~\eqref{AreaConstraint} reduces to $\int_{a_-}^{a_+} \tilde{q}\,\mathrm{d}x=1$. In the special case when $\tilde{q}=h/v$, namely
\begin{equation}\label{ParabolicFlux}
	q=\dot{v}h/v,
\end{equation}
the integrals $I_\pm$ evaluate to $I_\pm=\pm 3\dot{v}/(2d)$ at the orders we retain. In this case, the last two terms in~\eqref{kappa} cancel each other out, thus reducing Eq.~\eqref{IDESystem} to the same system of ordinary differential equations (ODEs) as in Ref.~\cite{Vellingiri2011}, with mass transfer effects entering only through the dependence of $\vartheta$ on $v$, see Eq.~\eqref{ApparentContactAngle}. Although, as far as we are aware, this particular form for $q$ does not correspond to some physically relevant scenario, some of the calculations that follow in the next section prescribe $q$ according to Eq.~\eqref{ParabolicFlux} to highlight a number of key features of the dynamics, which do not depend on the particular choice for $q$. The corresponding numerical solution to the PDE problem \eqref{full} when $q$ is given by Eq.~\eqref{ParabolicFlux} also avoids the need to use dense meshes in the interior of the domain that would be required to accurately capture highly localized fluxes.

We conclude the derivation of the reduced system~\eqref{IDESystem} by acknowledging some caveats surrounding a number subtle features of the present analysis. The first is that it relies on taking $|\dot{a}_\pm|\gg\lambda$, which is clearly violated as mass transfer switches from inflow to outflow and vice versa, as it may cause a droplet front to momentarily stop moving while switching its direction of motion. This issue however is of very brief duration and seldom gives noticeable departures from the solutions to the full problem \eqref{full}. The second caveat is that the dynamics for $t=O(1)$, during which the free surface of the droplet evolves towards its quasistatic shape, is not properly accounted for. This limit is not analytically tractable and mostly requires a numerical treatment (see, e.g.\ Ref.~\cite{Saxton2016} for the case of evaporating droplets). Just as in the first caveat, this relaxation towards quasisteady dynamics occurs in a short period without impacting the dynamics appreciably (see also also the beginning of Sec.~\ref{Sec:Results} for further remarks). The final caveat is that a more complete asymptotic procedure possibly requires a separate treatment for a receding droplet fronts, following, for example, the analysis by Eggers for the case of a receding contact line on a plate withdrawn from a liquid bath \cite{Eggers2005}. Given that receding fronts typically attain lower speeds than advancing ones, we chose not to pursue such analysis because these effects manifest themselves strongly only for sufficiently high recession speeds. Thus to extend the applicability of the asymptotic analysis and overcome the above-mentioned limitations, the development of a composite expansion would be required as a means to encapsulate all the pertinent scales present in the problem. This, however, appears to be a formidable task, the undertaking of which can be deemed unnecessary given the generally excellent agreement we observe between solutions of \eqref{full} and the asymptotic models, which is highlighted in the following section.

\section{Results}\label{Sec:Results}

\begin{figure}
	\includegraphics[scale = 1]{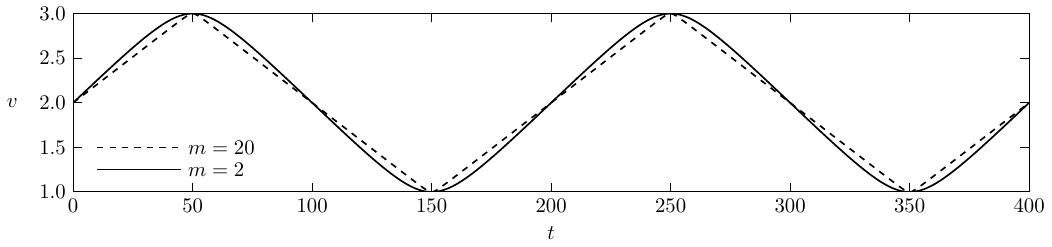}
	\caption{Sample evolutions for $v(t)$ according to Eq.~\eqref{AreaFunc} for $\bar{v}=2$, $\tilde{v}=1$ and $p=200$, when $m=2$ and $m=20$.}
	\label{Fig:AreaFunc}
\end{figure}

In this section, we offer a qualitative assessment of how well the predictions of the reduced system \eqref{IDESystem} compare with those of the full problem \eqref{full}. To ensure the efficient and accurate numerical solution to the full problem, we follow the methodologies developed by Savva and co-workers (see, e.g., Refs.~\cite{Savva2009,Savva2017}), namely we use the mapping~\eqref{CoordMap} and spectral collocation to discretize the interval $s\in[-1,1]$, in order to densely cluster points near the endpoints for resolving the boundary layers in $\pd[h]{x}$. In this manner, considerably fewer collocation points are required compared to, say, a less accurate finite differences scheme on a uniform mesh. Time stepping is performed with the method of lines, using MATLAB's \texttt{ode15s} stiff ODE solver. For the initial droplet shape, we chose profiles for given $a_\pm(0)$ and $v(0)$ that satisfy conditions \eqref{Cond1} and \eqref{Cond2} and closely match the leading-order profile in the outer region~\eqref{LeadingOrderHeight} (see also Ref.~\cite{Savva2017}). By doing so, the transients leading to quasistatic dynamics are significantly reduced, noting that the relaxation to quasistatic dynamics even for appreciably distorted initial profiles typically occurs within $t=O(10^{-2})$ during which the droplet fronts barely move from their initial position. Simulations with the reduced system~\eqref{IDESystem} can be carried out with any standard ODE solver, where the integrals $I_\pm$, Eq.~\eqref{Iintegral}, are evaluated with the Legendre--Gauss quadrature to avoid evaluation of the integrand at $s=\pm 1$ (see, e.g.\ \S25.4  in Ref.~\cite{AbramSteg1972}). Unless stated otherwise in the results that follow, we fix $a_\pm(0)=\pm 1$, $\lambda=10^{-4}$ and plot the solutions to the PDE and reduced systems with solid and dashed lines, respectively [see Ref.~\cite{repo} for a Python implementation of the reduced system, Eq.~\eqref{IDESystem}).

Whenever periodic variations of the cross sectional area are required, these are prescribed with the $p$-periodic function
\begin{equation}\label{AreaFunc}
v(t) = \bar{v}+\frac{\tilde{v}}{\arctan m}\arctan \left[\frac{m\sin(2\pi t/p)}{\sqrt{1+m^2\cos^2(2\pi t/p)}}\right],
\end{equation}
which describes oscillations of amplitude $\tilde{v}$ away from the mean value, $\bar{v}$. For finite values of the parameter $m$, \eqref{AreaFunc} is everywhere smooth, tending to $\bar{v}+\tilde{v}\sin(2\pi t/p)$ in the limit $m\to 0$. In the opposite limit, as $m\to\infty$, $v(t)$ tends to a piecewise linear $p$-periodic sawtooth function. Although the jump discontinuities in the first derivative of a sawtooth function are generally harmless and do not pose additional challenges in simulations, Eq.~\eqref{AreaFunc} avoids these discontinuities altogether while simulating nearly piecewise linear inflow/outflow scenarios for sufficiently large $m$. In all simulations performed, we fix $m=20$ (see Fig.~\ref{Fig:AreaFunc}).

\subsection{Hysteresis-like effects}
\begin{figure}[t!]
\includegraphics[scale = 1]{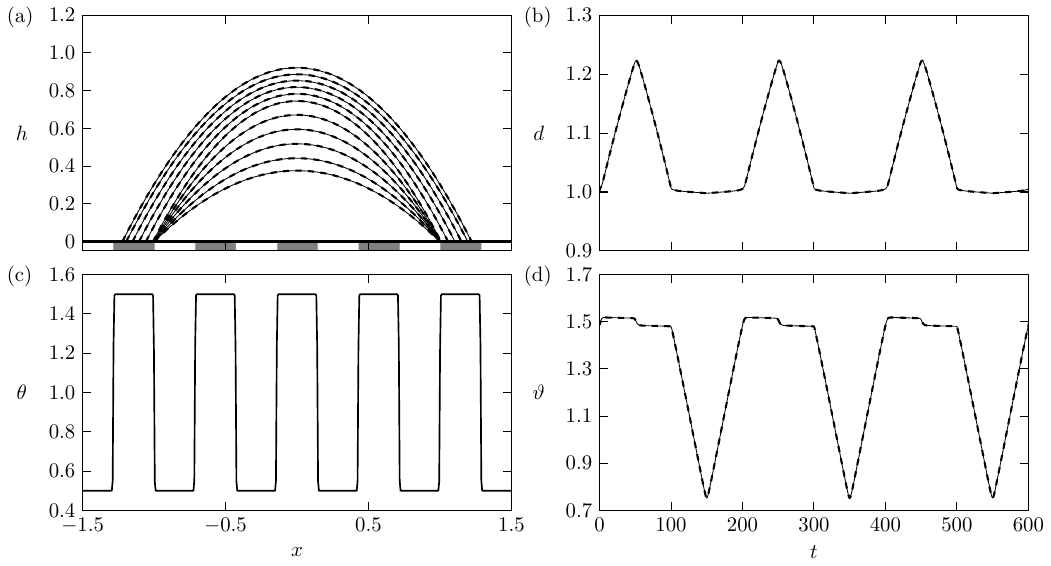}\caption{Alternating constant-angle and constant-radius modes on a substrate of alternating wettability patches with profile $\theta(x) = 1+0.5\tanh\left[30\cos(11x)\right]$. The mass flux is prescribed by Eq.~\eqref{ParabolicFlux} and $v(t)$ by \eqref{AreaFunc} with $\bar{v}=1$, $\tilde{v}=0.5$ and $p=200$. (a) Droplet profiles from $t=50$ to $t=150$ during mass loss in increments of $10$ time units. (b) and (d) evolution of the droplet half-width and apparent contact angle, respectively. The substrate in (a) is shaded according to the values of $\theta(x)$ (plotted in (c)), where $\theta\approx 1\pm 0.5$ are represented by the dark- and light-shaded patches, respectively.}
\label{Fig:ModeSwitch}
\end{figure}

\begin{figure}[t!]
\includegraphics[scale = 1]{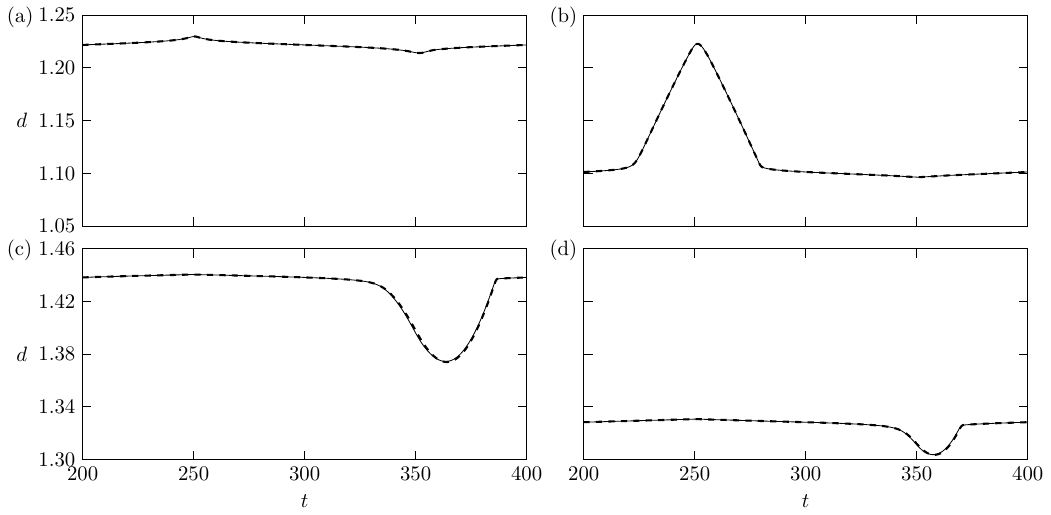}
\caption{Evolution of the half-width for the heterogeneity  profile  $\theta(x) = 1+0.5\tanh\left[30\cos(\alpha x)\right]$ and different values of $\alpha$ with all other parameters as in Fig.~\ref{Fig:ModeSwitch}. (a) $\alpha=9$; (b) $\alpha=10$; (c) $\alpha=12$; (d) $\alpha=13$. Compare with Fig.~\ref{Fig:ModeSwitch}(b) for which $\alpha=11$.}
\label{Fig:ModeSwitch2}
\end{figure}

\begin{figure}[t!]
	\includegraphics[scale=1]{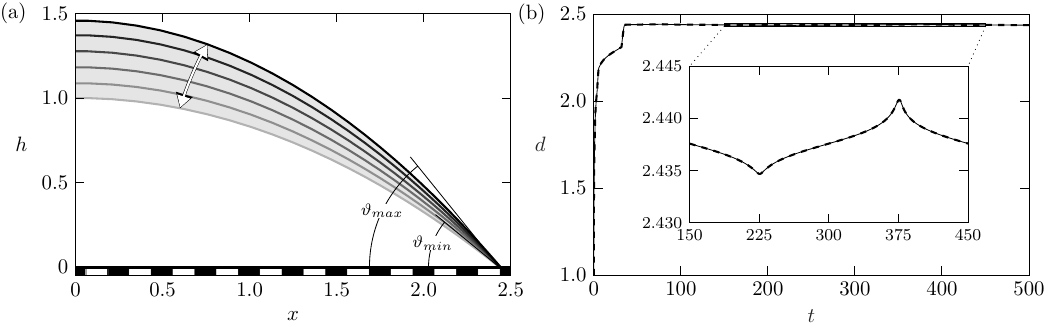}
	\caption{Macroscopic pinning on a substrate with profile $\theta(x) = 1+0.2\tanh\left[20\cos(8\pi x)\right]$. The mass flux is prescribed by Eq.~\eqref{ParabolicFlux} and $v(t)$ by Eq.~\eqref{AreaFunc} with $\bar{v}=4$, $\tilde{v}=0.75$ and $p=300$. (a) Droplet profiles separated by $25$ time units, over a period of liquid inflow/outflow ($\vartheta_{max}\approx1.2$; $\vartheta_{min}\approx0.8$). (b) Evolution of the droplet half-width, in which the contact points appears stationary macroscopically, but undergoes movements at the microscale (see inset).}
	\label{Fig:SlightVariation}
\end{figure}

In the first set of examples we assume that $q$ is given by Eq.~\eqref{ParabolicFlux} which, as previously mentioned, reduces Eq.~\eqref{IDESystem} to a simpler system of ODEs. In these cases we consider heterogeneity profiles that describe alternating patches of nearly constant wettability, in order to highlight some complex behaviors that arise. These include the pinning of the droplet fronts, which typically occurs in regions where $\theta(x)$ changes abruptly and points to substrate-induced hysteresis-like effects, as well as the constant-radius and constant-angle modes, previously discussed in Sec.~\ref{Sec:Introduction}. Indeed, for area changes following Eq.~\eqref{AreaFunc} and suitably chosen parameters $\bar{v}$, $\tilde{v}$, $p$ and profiles $\theta(x)$, the dynamics can be made to alternate between these two modes without treating both stages separately, as shown in Fig.~\ref{Fig:ModeSwitch}, which also highlights the excellent agreement between the full \eqref{full} and reduced systems \eqref{IDESystem}. Also noteworthy here is that similar observations were also reported by Pham and Kumar when considering numerically long-wave models of droplets in the presence of evaporative effects and a topographical defect for the cases of axisymmetric drops on impermeable surfaces~\cite{Pham2017} and 2D colloidal suspensions on permeable surfaces~\cite{Pham2019}.

It is important to emphasize that this behavior is not particularly robust and is sensitive to the system parameters. To demonstrate this, the calculation of Fig.~\ref{Fig:ModeSwitch} is repeated in Fig.~\ref{Fig:ModeSwitch2}, only changing the wavenumber of the heterogeneities and showing the evolution of the half-width $d(t)$ over a period after the fronts appear to settle to a periodic state. We observe that a different wavenumber can ultimately detune the dynamics of $d(t)$ shown in Fig.~\ref{Fig:ModeSwitch}(b) and the system is able to exhibit markedly different behaviors (from nearly complete pinning, Fig.~\ref{Fig:ModeSwitch2}(a), to enhanced spreading, Fig.~\ref{Fig:ModeSwitch2}(c)). However, in principle, one can retune to the same qualitative dynamics by modifying other system parameters accordingly, e.g.\ the manner mass transfer occurs.

A manifestation of contact angle hysteresis in experiments is that the fronts remain pinned whenever the apparent contact angle falls between the values of the so-called receding and advancing contact angles~\cite{Bonn2009}. The combined effects of chemical heterogeneities and mass transfer are able to capture such behaviors as well, without imposing \emph{a priori} contact angle hysteresis (see also Ref.~\cite{Schwartz1998}). This is demonstrated in Fig.~\ref{Fig:SlightVariation} for a substrate of alternating wettability patches. The droplet fronts appear to be macroscopically pinned where the static angle transitions abruptly between the minimum and maximum contact angles, with the apparent contact angle varying approximately between these values. However, by zooming into the evolution of the half-width we observe that the fronts always exhibit movement, albeit at the microscale (for a millimeter-sized droplet, this would correspond to sub-micrometer motions along the substrate), a feature which is very accurately captured by the lower-dimensional system~\eqref{IDESystem} (see inset of Fig.~\ref{Fig:SlightVariation}(b)). This use of heterogeneities as a plausible mechanism for hysteresis has also been invoked in the context of droplet motion on inclined surfaces to explain the pinning of the fronts as the inclination angle increased and the existence of a critical angle beyond which the substrate can no longer support the droplet at equilibrium~\cite{Savva2013}.

\subsection{Localized mass flux}

The simple flux distribution~\eqref{ParabolicFlux} used previously revealed some of the qualitative features of the dynamics. For arbitrary spatiotemporal flux variations, the motion of the moving fronts is captured by Eq.~\eqref{IDESystem} through the presence of $I_\pm$ terms, Eq.~\eqref{Iintegral}. In this section, the effects of localized mass transfer are explored, by representing $q$ with the scaled Gaussian
\begin{equation}\label{LocalMassFlux}
	q = \frac{2\dot{v}\sqrt{S}\e^{-S(x-x_0)^2}}{\sqrt{\pi}\left\{\text{erf}\left[\sqrt{S}\left(a_+-x_0\right)\right]-\text{erf}\left[\sqrt{S}\left(a_--x_0\right)\right]\right\}},
\end{equation}
whose peak is located inside the droplet footprint at $x_0$ and the prefactor is chosen to satisfy Eq.~\eqref{AreaConstraint} (here `erf' denotes the error function). When $S>0$ is sufficiently large, we can use Eq.~\eqref{LocalMassFlux} with $a_-(t)<x_0<a_+(t)$ to argue that $q_\pm\approx0$, so that we may use the reduced system~\eqref{IDESystem} instead of the more complete system~\eqref{transcendental}. As mentioned previously, the full problem~\eqref{full} requires a considerably denser mesh to resolve the spatial variations of $q$ prescribed according to Eq.~\eqref{LocalMassFlux}. For this reason, we used only moderate values for $S$ to achieve a satisfactory compromise between a fast decay of $q$ towards zero, and computational efficiency.

\begin{figure}[t!]
	\includegraphics[scale = 1]{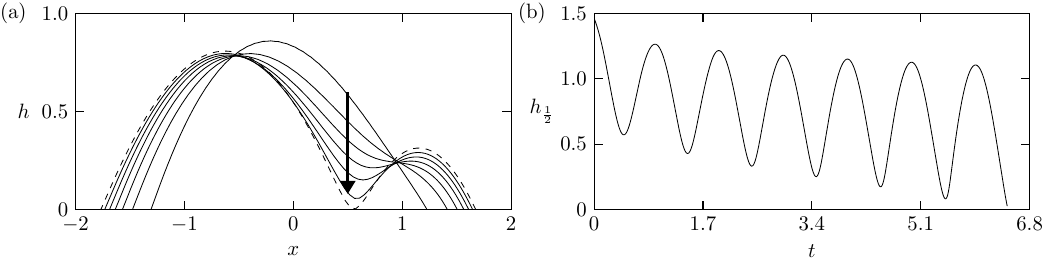}
	\caption{Droplet breakup due to a localized mass flux when  $\theta(x) = 1$. Here fluid is injected/removed at $x_0=0.5$ and $q$ is given by \eqref{LocalMassFlux} with $S=15$ and $v(t)=2+0.6\cos(2\pi t)$. (a) Droplet profiles from $t=0.5$ in increments of $1$ time unit until $t=5.5$ (direction of time indicated by the arrow). The dashed curve shows the profile when $t=6.45$ where the computation was terminated. (b) Evolution of the droplet height at $x=0.5$ showing the thinning of the profile where $q$ is localized.}
	\label{Fig:HighAdot}
\end{figure}

It is worth noting that localized fluxes may lead to droplet breakup when  $|\dot{v}|\gg|\dot{a}_\pm|$, which is beyond the quasistatic limit of applicability of our theory. In such cases, the full PDE needs to be considered. Such an example is shown in Fig.\ \ref{Fig:HighAdot} where we observe that the rapidly changing localized flux creates a neck region in the vicinity of the fluid inlet/outlet, which progressively becomes thinner and ultimately leads to breakup. In this example, the calculation was terminated just before the droplet height vanished at some point between the two contact points, thus avoiding the development of schemes to deal with the actual breakup and the resulting dynamics after it occurs.

\begin{figure}[t!]
	\includegraphics[scale = 1]{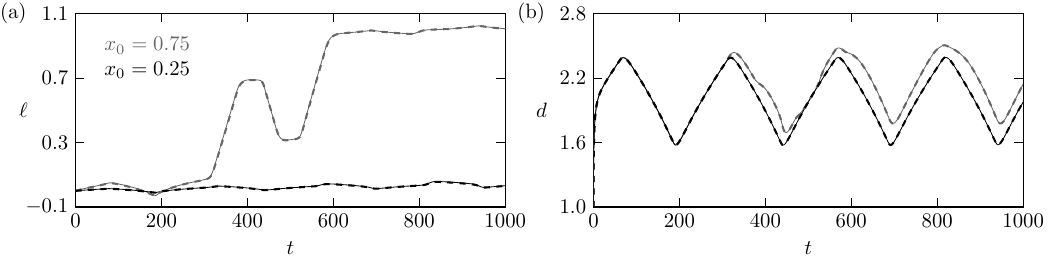}
	\caption{Effect of changing the position of the flux in over the substrate with profile $\theta(x) = 1+0.1\tanh\left[5\cos(\pi x)\right]$. The mass flux is given by Eq.~\eqref{LocalMassFlux} with $S=20$ and $v(t)$ is prescribed by Eq.~\eqref{AreaFunc} with $\bar{v}=3$, $\tilde{v}=1.25$ and $p=250$. (a) and (b) show the evolution of the droplet midpoint and half-width for differently positioned fluxes.}
	\label{Fig:LocalisedComparison}
\end{figure}

Many of the interesting features reported previously for fluxes of the form \eqref{ParabolicFlux} also pertain for the localized fluxes as well. However, if the mass flux is localized, more control may be exercised on how the droplets move on surfaces. If mass transfer occurs sufficiently slowly, it causes the droplet to move, to the extent permitted by heterogeneities, so as to center around the inlet/outlet point. This, however, may require more time for the droplet to settle to periodic motion. This effect is highlighted in  Fig.~\ref{Fig:LocalisedComparison}, showing that the droplet midpoint defined as $\ell = (a_++a_-)/2$ evolves very differently depending on where the inlet/outlet is located. Importantly, the excellent agreement of the reduced model with the full equations demonstrates the importance of including the $O(\dot{v}/|\ln \lambda|)$ terms in Eq.~\eqref{IDESystem}, which merely correspond to higher-order contributions to the leading-order $O(1/|\ln\lambda|)$ terms.

\begin{figure}[t!]
	\includegraphics[scale = 1]{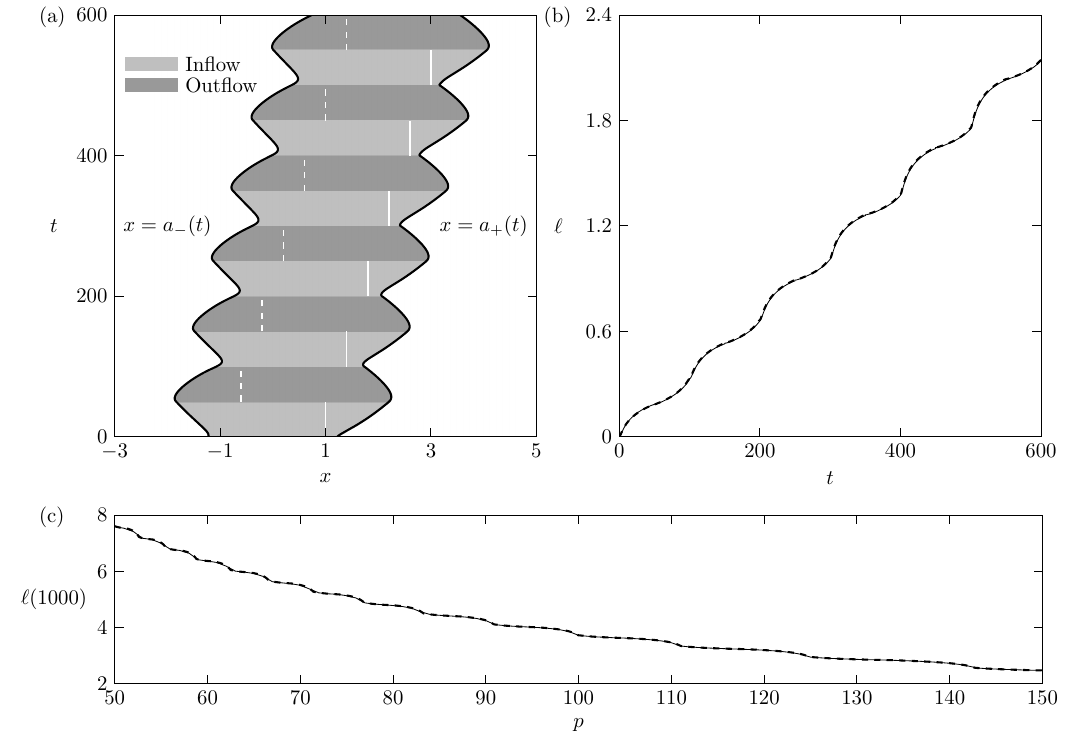}
	\caption{Droplet transport using the delta function flux $q=\dot{v}\delta(x-x_{0}(t))$ over the homogeneous substrate $\theta(x) = 1$, where $x_{0}(t)$ is given by Eq.~\eqref{eq:switch}. The area fluctuations  using Eq.~\eqref{AreaFunc} with $\bar{v}=2$, $\tilde{v}=1$ and $t$ replaced by $t-p/4$. (a) Space--time plot showing the droplet footprint when $p=100$, demarcated by the black solid curves $x=a_\pm(t)$. Dark/light patches correspond to fluid outflow/inflow stages. The solid and dashed white lines show the positions at which fluid is injected and removed, respectively. (b) the evolution of the droplet midpoint, $\ell(t)$ corresponding to the plot of (a). (c) $\ell(1000)$ as a function  of the flow period, $p$.}
	\label{Fig:Delta}
\end{figure}

In the limit $S\to\infty$, Eq.~\eqref{LocalMassFlux} reduces to the a Dirac delta function distribution, namely $q = \dot{v}\delta(x-x_0)$, which allows us to evaluate the integral $I_\pm$ in  Eq.~\eqref{Iintegral} giving
\begin{equation}
	I_\pm = \dot{v}\left[\frac{1}{2d}\ln\left(\frac{x_0-a_-}{a_+-x_0}\right)+\frac{1}{a_\pm-x_0}\right],
\end{equation}
that is valid when $x_0\neq a_{\pm}(t)$. Equations~\eqref{IDESystem} reduce to a system of ordinary differential equations, whose implementation is rather straightforward; for the full equations, we consider numerically an integrated form of Eq.~\eqref{GovPDE} over the spatial variable. Figure~\ref{Fig:Delta} shows a sample calculation in the case when $\theta(x)=1$. Here the droplet is initially at an equilibrium corresponding to $v(0)=1$, $a_\pm(0) = \pm\sqrt{3/2}$. We periodically inject/extract fluid with period $p$ using Eq.~\eqref{AreaFunc} with $t$ replaced by $t-p/4$. During the first half of the period we inject fluid, which is then removed during the second half of the period. At the same time, we move the corresponding injection/extraction point in such a manner to induce directed droplet transport to the right. Specifically, during the $n$-th cycle, $n\ge1$, we use $q = \dot{v}\delta(x-x_{0}(t))$, where sources/sinks are activated according to 
\begin{align}
x_{0}(t) = \begin{cases}
\dfrac{3}{5} + \dfrac{2n}{5} & (n-1)p \le t< (n-1/2)p\quad\text{(source)}\\[10pt]
-1+\dfrac{2n}{5} & (n-1/2)p \le t < np\quad\text{(sink)}
\end{cases}.\label{eq:switch}
\end{align}
This means that injection/extraction points are located at $x=2k/5-1$, where $k$ is a positive integer (the dynamics show that all points with $k\ge5$ can act both as sources and sinks). The way transport is induced is by injecting fluid from a point close to the right contact point and removing it from a point close to the left contact point. It turns out that the efficiency of droplet transport, as measured in terms of the distance travelled for fixed $p$, increases as the injection/extraction points tend towards the right/left contact points. However, this violates our original assertion that there is no mass flux through the contact lines. Although a more detailed parametric investigation is beyond the scope of the present work, it is worth remarking that an alternative approach to promote transport is by lowering the period $p$ (see Fig.~\ref{Fig:Delta}c).

\subsection{Transition to periodic dynamics}

The previous examples demonstrate that the dynamics becomes periodic in the long-time limit if mass changes are also periodic. Apart from the cases shown in Fig.~\ref{Fig:LocalisedComparison}, we observe that the transition to periodic motion occurs around the first two periods of inflow and outflow, but it may be significantly prolonged depending on the heterogeneity profile. When $\theta(x)$ is suitably chosen it may allow for longer excursions away from the initial droplet position, which may violate the assumption for the flux to be localized somewhere between the two contact points for Eq.~\eqref{IDESystem} to hold. A remedy for this issue is to allow the location of the inlet/outlet to follow the droplet as time progresses. However, for studying these transitions, it suffices to just represent $q$ by  Eq.~\eqref{ParabolicFlux}.

\begin{figure}[t!]
\includegraphics[scale = 1]{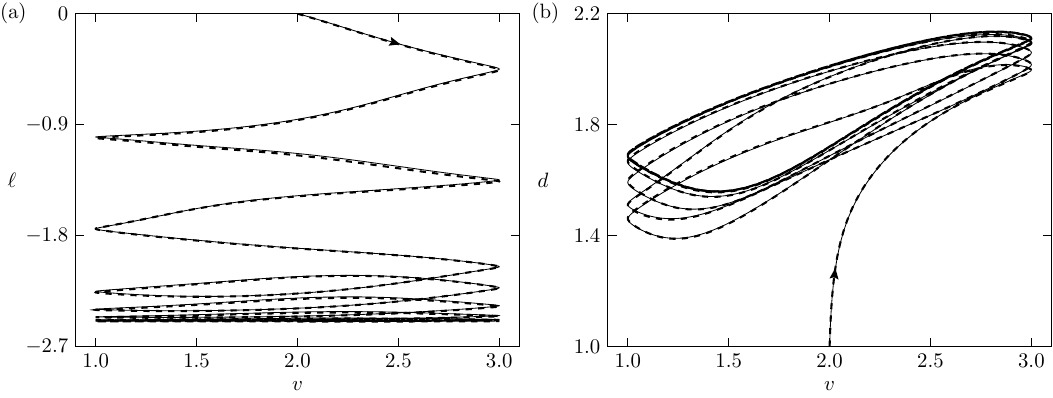}
\caption{Slow settling to periodic dynamics when  $\theta(x)= 1+0.1\cos(8\pi x/5)+0.2\sin(\pi x/5)$. The mass flux is prescribed by Eq.~\eqref{ParabolicFlux} and $v(t)$ by Eq.~\eqref{AreaFunc} with  $\bar{v} = 2$, $\tilde{v}=1$ and $p = 100$. (a) and (b) midpoint $\ell$ and half-width $d$ as functions of the droplet area $v$, respectively.}
\label{Fig:Varied1}
\end{figure}

Such a calculation is depicted in Fig.~\ref{Fig:Varied1}, where a shorter period of mass transfer and a combination of harmonics for $\theta(x)$ is used, plotting how the half-width $d$ and midpoint $\ell$ evolve with the droplet area $v(t)$. We readily observe that the droplet requires several more cycles to settle to a periodic motion, where, once more, the reduced model \eqref{IDESystem} excellently captures the dynamics predicted by the full model \eqref{full}. For this choice of $\theta(x)$ the wettability contrasts are not as pronounced compared to, say, those of Fig.~\ref{Fig:ModeSwitch}, which allows the droplet fronts to move more freely. As a result, the pinning effects are weaker, which appear to prolong the transition to periodic dynamics.

\begin{figure}[t!]
	\includegraphics[scale = 1]{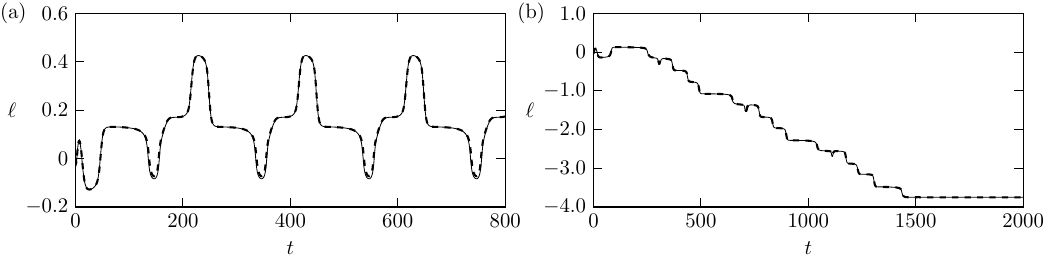}
	\caption{Evolution of the droplet midpoint, $\ell$, on a heterogeneous surface with $\theta(x) = 1+0.1\cos(8\pi x/5)+0.3\sin(10\pi x/3)$, where $q$ is given by  Eq.~\eqref{ParabolicFlux} and $v$ by Eq.~\eqref{AreaFunc}, with $\bar{v}=2.5$, $\tilde{v} = 1.5$ when (a) $p=200$ and (b) $p=400$. The larger value of $p$ exhibits a delayed transition to periodic motion.}
	\label{Fig:Varied2}
\end{figure}

Hence, it is important to acknowledge that the dynamics is strongly influenced by the interplay between mass flux and substrate features. In Fig.~\ref{Fig:Varied2}, we compare the mid-point dynamics for two different periods of inflow/outflow using Eq.~\eqref{ParabolicFlux} for a droplet moving over a substrate with $\theta(x) = 1+0.1\cos(8\pi x/5)+0.3\sin(10\pi x/3)$. For the shorter period ($p=200$), the droplet settles to a periodic state within the first period of inflow/outflow (Fig.~\ref{Fig:Varied2}(a)); for the longer period ($p=400$) the droplet undergoes multiple mass flux cycles before settling to the periodic state, while it is being shifted an order unity distance to a different region of the substrate  (Fig.~\ref{Fig:Varied2}(b)). This example highlights that more slowly varying mass fluxes are more strongly influenced by substrate heterogeneities, thus suggesting that the effects of surface heterogeneities can be mitigated by increasing the mass flux rates, and that we can afford more control on droplet motion by coupling slowly varying fluxes with appropriately tuned surface heterogeneities.

\subsection{Snapping droplets}

\begin{figure}[t!]
	\includegraphics[scale = 1]{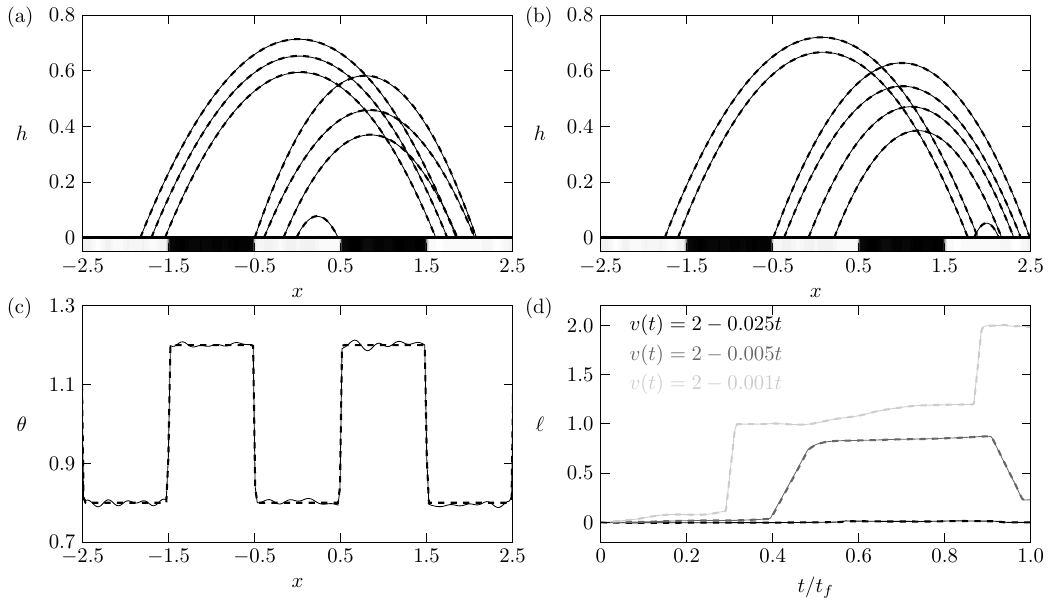}
	\caption{Snapping mode dynamics using Eq.~\eqref{ParabolicFlux} for the mass flux and linear mass loss according to $v(t)=2-wt$  (a) Droplet profiles when $w=0.005$ plotted at times $t=50$ to $t=300$ in increments of 50 time units with the last profile at $t=395$. (b) Profiles for area when $w= 0.001$ at times $t=250$ to $t=1500$ in increments of $250$ time units and the last profile at $t=1990$. (c) The realization of the heterogeneity profile which is used in simulations (solid curve) and its noise-free counterpart given by $\theta(x) = 1-0.2\tanh\left[50\cos(\pi x)\right]$ (dashed curve). (d) Evolution of the mid-point as a function of the normalized time $t/t_f$, where $t_f = 79$ for $w = 0.025$, $t_f = 399$ for $w = 0.005$, and $t_f = 1999$ for $w = 0.001$.}
	\label{Fig:Snapping}
\end{figure}

We have seen thus far that stick-slip events occur on typically fast time scales if the contact line is temporarily trapped on sharp wettability contrasts. In a recent study, Wells \emph{et al.}\ reported that for a sufficiently slowly evaporating droplet a different mode can be observed, which is marked by a series of distinct snapping events, during which the droplets shift towards a different location on the substrate~\cite{Wells2018}. Specifically, Wells \emph{et al.}\ considered sinusoidal surface topographies both experimentally and numerically, and observed that droplets can break the left-right symmetry as their contact line retracts due to evaporation, allowing them to move as a whole to a new location but at a slower timescale than typical stick-slip jumps. Similar effects may be observed with chemically heterogeneous surfaces as well. This is depicted in Fig.~\ref{Fig:Snapping} for droplets of linearly decreasing mass and a substrate with heterogeneities prescribed according to
\begin{equation}\label{SnappingHet}
	\theta(x) = 1-\frac{1}{5}\tanh\left[50\cos(\pi x)\right]+\tilde{\theta}(x),
\end{equation}
where $\tilde{\theta}(x)$ corresponds to band limited white noise represented by a superposition of harmonics with wavenumbers up to $10\pi$, whose amplitudes are normally distributed with zero mean and standard deviation 0.005. This profile has a lower wettability contrast to that of Fig.~\ref{Fig:ModeSwitch}, whereas the low amplitude random features introduced by $\tilde{\theta}(x)$ promote the breaking of symmetry (see Fig.~\ref{Fig:Snapping}(c)). A comparison of Figs.~\ref{Fig:Snapping}(a) and \ref{Fig:Snapping}(b) shows that the rate at which mass is lost plays a key role in the underlying motion. In both cases we initially have symmetrically receding fronts, until a snapping event occurs which shifts both droplets to the right. Depending on the rate of mass loss, Fig.\ \ref{Fig:Snapping} shows that other snapping events may occur prior to extinction, pushing the droplet to the left (faster mass loss, Fig.~\ref{Fig:Snapping}(a)) or to the right (slower mass loss, Fig.~\ref{Fig:Snapping}(b)). Figure~\ref{Fig:Snapping}(d) shows the evolution of the droplet mid-point in normalized time units and clearly demonstrates that the faster rate of mass loss completely suppresses snapping events.  This further corroborates our earlier assertion that, for heterogeneities to play a more predominant role, the rate at which fluid inflow/outflow occurs needs to be sufficiently slow. Similar observations can also be made with dynamics of mass gain as well, or even if mass transfer occurs periodically.

\section{Bifurcation structure}\label{sec:bifurcations}

\begin{figure}[t!]
	\includegraphics[scale=1]{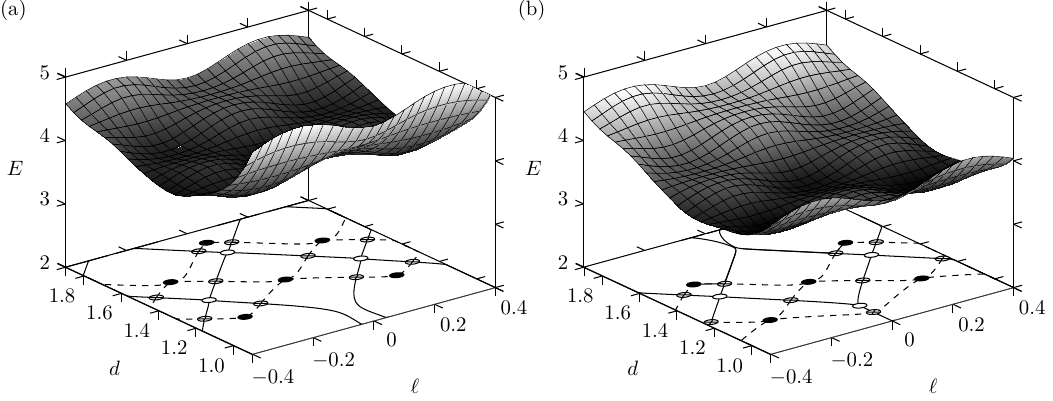}
	\caption{Interfacial energy for surfaces with heterogeneities prescribed according to $\theta(x) = 1+0.5\cos(4\pi x)$ when (a) $v=1$, and (b) $v=1.25$. Solid, open and crossed circles correspond to the stable, unstable and saddle points, respectively. Solid and dashed lines show the stable and unstable manifolds for the saddle nodes, respectively.}
	\label{fig:EnergyPlots}
\end{figure}

To better understand the mechanisms for the behaviors observed as the system parameters vary, an alternative approach involves investigating the bifurcation structure of the equilibria of the system for fixed values of $v$. Hence, by treating $v$ as a bifurcation parameter, we trace how the location, nature and stability of these equilibria vary, which can then be contrasted with time-dependent simulations. For fixed $v$, the equilibria and their nature can be determined from the interfacial energy of the system, defined in dimensionless long-wave form as (see Ref.~\cite{Vellingiri2011})
\begin{equation}
	E(\ell,d) = \integ{\ell-d}{\ell+d}{\left[\left(\pd[h_0]{x}\right)^2+\theta(x)^2\right]}{x}.
\end{equation}
The extrema of $E$ correspond to the equilibria of the system, whose nature and stability can be assessed straightforwardly from the determinant of the Hessian matrix of $E(\ell,d)$. Figure\ \ref{fig:EnergyPlots} shows such an example for a specific heterogeneity profile and two nearby values of $v$. The figure also shows the equilibria projected on the $\ell-d$ plane together with the stable and unstable manifolds of the saddle points, which were computed from the IDE system~\eqref{IDESystem} for $q\equiv0$ and demarcate the basins of attraction of the stable equilibria. We see that changes in $v$ result into bifurcations that alter the structure of the corresponding phase plane. By comparing Figs.~\ref{fig:EnergyPlots}(a) and \ref{fig:EnergyPlots}(b), we observe the formation and destruction of equilibria, which ultimately reveal that dynamic changes in $v$ can lead to appreciable changes in the motion of the fronts. For example, compare the two phase portraits in Fig.~\ref{fig:EnergyPlots} in the close vicinity of the line $\ell=0$; equilibria disappear when they collide as $v$ is increased (for $d\approx1.7$), or new equilibria form when previously non-intersecting manifolds meet (when $d\approx1$).  When such topological changes/bifurcations occur for sufficiently slow variations in $v$, the droplet responds to these changes by appreciably altering its motion. In other words, the droplet fronts are expected to closely trace how the corresponding equilibria evolve had $v$ been treated as a bifurcation parameter, and the observed de-pinning events, such as stick-slip, the snapping mode or the transition from the constant-radius to the constant-angle modes are merely manifestations of the dynamics that arise due to the topological changes in the  basins of attraction of nearby equilibria.

\begin{figure}[t!]
	\includegraphics[scale = 1]{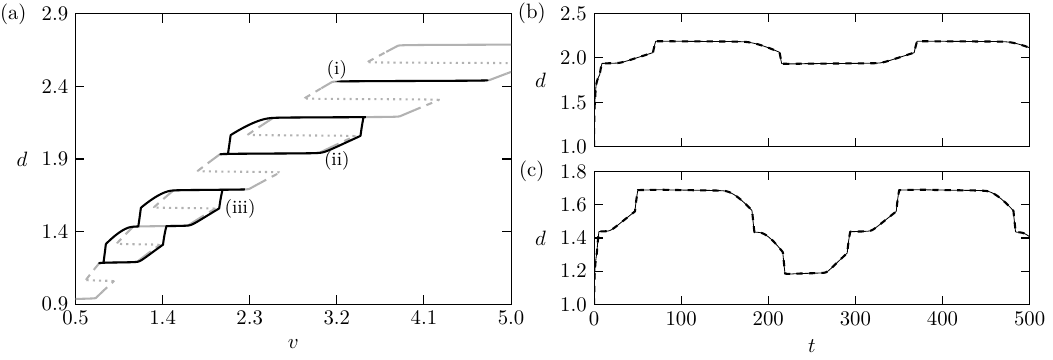}
	\caption{(a) Bifurcation diagram projected on the $v-d$ plane when $\theta(x)=1+0.2\tanh\left[20\cos(8\pi x)\right]$ with overlaid dynamic simulations. Gray solid, dashed and dotted curves correspond to stable, saddle and unstable branches of the bifurcation diagram, respectively. Black curves are solutions to the reduced system \eqref{IDESystem}, obtained with the same parameters as in Fig.~\ref{Fig:SlightVariation} but with different values for $\bar{v}$: (i) $\bar{v}=4$; (same as in Fig.~\ref{Fig:SlightVariation}) (ii) $\bar{v}=2.75$ and (iii) $\bar{v}=1.5$. (b) and (c) evolution of the droplet half-width corresponding to curves (ii) and (iii) in (a), respectively, as obtained from the full \eqref{full} (solid curves) and reduced \eqref{IDESystem} systems (dashed curves).}
	\label{Fig:TanhBif}
\end{figure}

\begin{figure}[t!]
	\includegraphics[scale = 1]{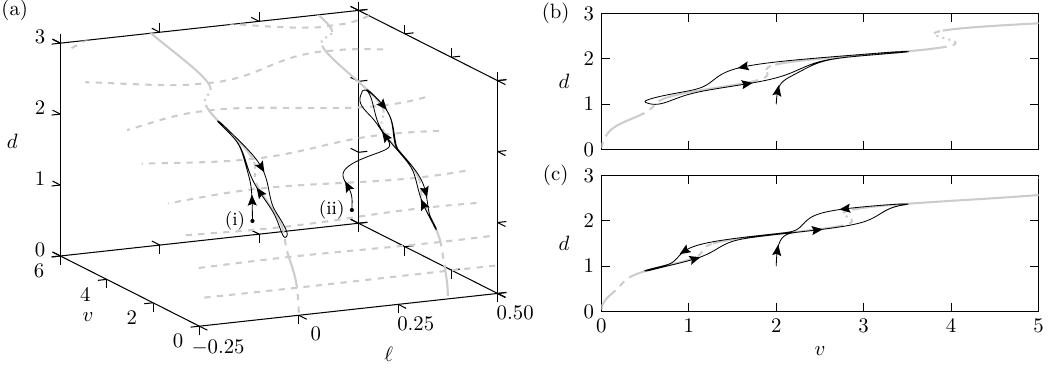}
	\caption{(a) Bifurcation structure for the substrate defined by $\theta(x)=1+0.15\cos(8\pi x/3)$ in the $v-\ell-d$ plane (gray curves) with two overlaid calculations of the reduced system (black curves labelled (i) and (ii)). The styles of the gray curves are as in Fig.~\ref{Fig:TanhBif}(a). Curves (i) and (ii) are obtained from Eqs.~\eqref{IDESystem} and \eqref{ParabolicFlux}; $v$ is given by Eq.~\eqref{AreaFunc} with $\bar{v}=2$, $\tilde{v}=1.5$ and $p=250$, using $a_\pm(0) = \pm 1$ for curve (i) and  $a_+(0)=0.5$ and $a_-(0)=-1.5$ for curve (ii). (b) and (c) slices of the bifurcation diagram in (a) showing, respectively, curves (i) and (ii) projected on  the $v-d$ plane.}
	\label{Fig:SingleHarmonicBif}
\end{figure}

\begin{figure}[t!]
	\includegraphics[scale = 1]{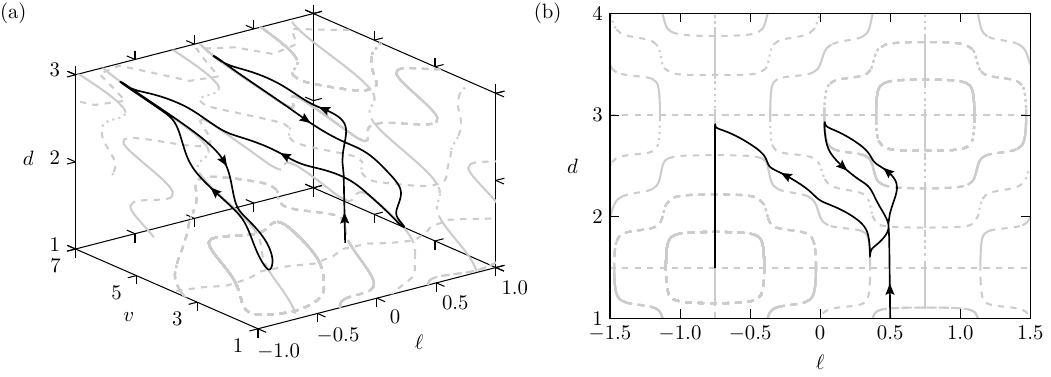}
	\caption{Delayed transition to periodic dynamics for the substrate $\theta(x) = 1+0.15\cos(8\pi x/3)+0.05\sin(2\pi x)$ as visualized on a bifurcation diagram. (a) Bifurcation structure in the $v-\ell-d$ plane with an overlaid ODE trajectory obtained from Eqs.~\eqref{IDESystem} and \eqref{ParabolicFlux}; $v$ is given by Eq.~\eqref{AreaFunc} with $\bar{v}=4$, $\tilde{v}=2.5$ and $p=200$, using $a_+(0) = 1.5$ and $a_-(0) = -0.5$. (b) Projection of the plots in (a) on the $\ell-d$ plane. The style of the various curves are as in Fig.~\ref{Fig:TanhBif}.}
	\label{Fig:MultipleHarmonics}
\end{figure}

Similar work has been undertaken by Pradas \emph{et al.}\ using diffuse interface simulations in 2D~\cite{Pradas2016}. Unlike the work in Ref.~\cite{Pradas2016}, our work is limited to small contact angles, but our asymptotic analysis allows for a more efficient exploration of the parameter space, as well as a longer simulation of the system over many cycles of mass gain and loss to capture the transition to periodic motion. The remainder of this section is devoted to uncovering these bifurcation structures in the 3D space spanned by $\ell$, $d$ and $v$. The equilibrium branches were obtained by pseudo arc-length numerical continuation techniques for carefully chosen conditions to trace each branch~\cite{Allgower1990}.

One of the most interesting features shown in this work, which is not discussed in Ref.~\cite{Pradas2016} and is worthy of further investigation is the apparent contact angle hysteresis in Fig.~\ref{Fig:SlightVariation}, namely that the droplet appears to be pinned as liquid is pumped in and out of the droplet, but moves at smaller spatial scales. Figure~\ref{Fig:TanhBif}(a) shows an overlay of the data in Fig.~\ref{Fig:SlightVariation} (plot (i)) and the bifurcation diagram projected on the $v-d$ plane. For this example, $v$ varies between $3.25$ and $4.75$, and is able to trace one of the stable branches of the bifurcation diagram for which $d$ is nearly constant. Had $v$ oscillated with a sufficiently larger amplitude beyond the span of the stable branch, de-pinning of the fronts and a transition to another stable branch with different $d$ would have occurred. This is shown by plotting the evolution of the half-width  for droplets of smaller mean size in Fig.~\ref{Fig:TanhBif}(b) (plot (ii) in (a)) and Fig.~\ref{Fig:TanhBif}(c) (plot (iii) in (a)), where the corresponding pinning/de-pinning events are clearly distinguished.

When substrates are decorated with strong wettability contrasts, the dynamics described above undergo abrupt transitions as the stable branches of the bifurcation diagram are traversed (see Fig.~\ref{Fig:TanhBif}). In reality, the moving fronts only remain close to the bifurcation branches, but do not exactly trace them. This is a consequence of the fact that the bifurcation diagrams describe droplet equilibria for fixed values of $v$, whereas simulations capture dynamic phenomena and it is the quasisteady character of the dynamics that allows us to make meaningful visual comparisons between dynamic behaviors and these diagrams. If the heterogeneity features are represented by a single harmonic, the resulting equilibrium branches will contain more readily identifiable pitchfork bifurcations, as shown in Fig.~\ref{Fig:SingleHarmonicBif}. Like before, the droplet dynamics closely follows the stable branches of the system away from bifurcation points. Close to a bifurcation point, where typically a stable branch becomes unstable as $v$ is varied, the dynamics of the  system is slower to respond to this change and tends to overshoot away from the bifurcation points before approaching a nearby stable branch shortly afterwards. This overshooting, which has also been observed in Ref.~\cite{Pradas2016}, and, more generally, the departure of the actual dynamics away from the bifurcation curves becomes more pronounced if the contact lines become more mobile, which may also result from more rapidly changing mass fluxes and stronger slip effects.

A more realistic representation of the heterogeneities of an actual substrate is expected to consist of many more harmonics than the simple profile considered in Fig.~\ref{Fig:SingleHarmonicBif} due to typically unavoidable randomness in its features~
\cite{Savva2010}. Such representations, however, make the investigation of the corresponding bifurcation structures unwieldy. To demonstrate this, Fig.~\ref{Fig:MultipleHarmonics} shows how the complexity of the bifurcation structure of Fig.~\ref{Fig:SingleHarmonicBif} increases with the inclusion of an additional, low-amplitude harmonic in the heterogeneity profile of Fig.~\ref{Fig:SingleHarmonicBif}. For such a profile and depending on the initial state of the system, we can anticipate delayed transitions to periodic motion as the droplet fronts navigate the space of nearby quasisteady states until they reach a limit cycle in the long-time limit (see, e.g., the overlaid trajectory in Fig.~\ref{Fig:MultipleHarmonics}).

\section{Concluding remarks}\label{Sec:Conclusion}

Understanding the effects of mass transfer on droplet spreading is of high importance for a variety of industrial applications, and, utilising predictive models to simulate contact line spreading can facilitate the development and improvement of these technologies, as well as enhance our understanding of the natural world. However, the study of the underlying dynamics is inherently complex due to the multi-scale nature of the problem. Therefore, in this exploratory work we considered the 2D geometry to elucidate some of the key observed behaviors, as well as to gain new perspectives and insights into unravelling the dynamics of the arguably more complex 3D problem. Specifically, the method of matched asymptotic analysis was leveraged to derive a set of simplified models which approximate the solution of the governing PDE in the limit of vanishingly small slip lengths and slow contact line velocities, which in all cases considered, predict the dynamics of the full problem with remarkable accuracy, at least within the domain of applicability of the theory.

The analysis undertaken holds for any spatiotemporal dependence for the flux term and pertains in the distinguished limit when $\dot{v}$ and $\dot{a}_\pm$ are both $O(1/|\ln \lambda|)$, so that changes occur sufficiently slowly. In this limit, a quasistatic expansion is applicable so that the contact line velocities and the mass transfer terms are considered in the next-to-leading-order terms in the asymptotic expansion, extending the study in Ref.~\cite{OLIVER2015}, which considered ideal surfaces, constant $q$ and retained only $O(1/|\ln\lambda|)$ terms in the asymptotics. Although the evolution equations we obtained apply for arbitrary mass fluxes, we opted to limit our discussion to cases where it vanishes at the contact points, primarily to avoid any implementation difficulties that would have arisen had we solved for the evolution of the droplet fronts using the transcendental equations~\eqref{transcendental}. In this particular limit, the simpler set of IDEs, Eqs.~\eqref{IDESystem}, was obtained, which can essentially be viewed as an augmented Cox--Voinov law that accounts for mass transfer effects.

Although these models are, strictly speaking, not valid at early times or when $\dot{a}_\pm\to0$, the results presented suggest that we may confidently use them at all times, without compromising the generally excellent agreement with the predictions of the full PDE. To investigate some of the generic features of the dynamics in further detail, we considered two kinds of distributions for the liquid flux, namely one that scales with the droplet thickness according to Eq.~\eqref{ParabolicFlux} and one that mimics a more localized flux distribution, as described by Eq.~\eqref{LocalMassFlux}. While the former does not correspond to a physically motivated flux distribution, it allows for a further simplification of the IDE system~\eqref{IDESystem} that is nearly identical to the system obtained for the case of constant mass~\cite{Vellingiri2011}. For more general flux distributions, the spatial dependence of $q$ enters Eq.~\eqref{IDESystem} through the integral terms, $I_\pm$. Although they appear as higher-order corrections in the analysis, their presence is needed in order to accurately capture the dynamics.

With a number of representative cases, we highlighted the intricate interplay among the various effects, demonstrating how droplet behavior can drastically change even if small changes are introduced to either the chemical heterogeneity or fluid flow properties. Some of the key contributions of the present work include complementing related works undertaken in 2D~\cite{Vellingiri2011,OLIVER2015,Pradas2016}, where we show that it is indeed possible to observe hysteresis-like effects without explicitly assuming \emph{a priori} the presence of hysteresis (see, e.g. Figs.~\ref{Fig:ModeSwitch} and \ref{Fig:SlightVariation}), as also observed in studies with mass transfer \cite{Schwartz1998,Pradas2016}, and without (see, e.g.\ \cite{Savva2009, Savva2013, Vellingiri2011}). Crucially, we have also shown that the \emph{apparent, macroscopic} pinning of the fronts is also possible in this setting, and examined how such effects manifest themselves on the phase portrait (compare, e.g. Figs~\ref{Fig:SlightVariation} and \ref{Fig:TanhBif}). Also noteworthy is that the various modes for evaporation, including the recently discovered snapping mode \cite{Wells2018}, and the constant-radius and constant-angle modes can occur naturally in a single simulation once chemical heterogeneities are present (see also Ref.~\cite{Pham2017} who study the constant-radius and constant-angle modes over a topographical defect). However, these modes can be suppressed or amplified by appropriately tuning the chemical heterogeneity (see Figs.\ \ref{Fig:ModeSwitch} and \ref{Fig:ModeSwitch2}) and/or  the manner mass transfer occurs (see Fig.\ \ref{Fig:Snapping}). In all simulations performed, periodic mass fluxes led to periodic dynamics in the long-time limit, but the time required for the fronts to settle to periodic motion is highly dependent on the structure of the heterogeneities as well as the choice of the initial conditions (see Fig.~\ref{Fig:Varied2}(b)).  

Although the outcomes of a 2D model cannot be straightforwardly scrutinized by experiments, the combined analytical and computational work we have undertaken made looking into the complex bifurcation structure of the dynamics possible; gaining further insights into how hysteresis-like effects and transients to periodic motion occur by following the topological changes that take place as the nature and stability of droplet equilibria evolve with changing the droplet area. Importantly, the outcomes of this work are to be combined with the recent developments presented by Savva \emph{et al.}\ on 3D droplet motion over chemically heterogeneous surfaces \cite{Savva2018}, to offer a novel extension to the fully 3D geometry, which is the subject of the second part of the present study~\cite{Savva2021}.

\section*{Acknowledgement}
DG acknowledges support from the Engineering and Physical Sciences Research Council of the UK through Grant No EP/P505453/1; NS acknowledges support from the European Union's Horizon 2020 research and innovation programme under grant agreement No 810660.

\appendix

\section{Calculation of $\beta_\pm$}\label{BetaCalc}

The $\beta_\pm$ terms required for the outer region are found from an analysis Eq.~\eqref{eq:h1}, and may be determined without solving for $h_1$ itself. Namely, we use the chain rule to write $\pd[h_0]{t}=\dot{a}_{+}\pd[h_0]{a_+}+\dot{a}_-\pd[h_0]{a_-}+\dot{v}\pd[h_0]{v}$, integrate Eq.~\eqref{eq:h1} once and after some term re-arrangement, we find
\begin{equation}\label{OR:h1eq}
	\pddd[h_1]{s}=\frac{d^4}{h_0^3}\integ{-1}{s}{q(\tilde{s},t)}{\tilde{s}}+\frac{d^4\vartheta}{4h_0^3}\left(f_+\dot{a}_+\! + \!f_-\dot{a}_-\right)+\frac{d^3\dot{v}}{4h_0^3}(s-2)(1+s)^2,
\end{equation}
where $f_\pm = (1\mp s)(1\pm s)^2$. Next, we multiply Eq.~\eqref{OR:h1eq} by $f_\pm$ and integrate with respect to $s$ over the interval $[-1+\varepsilon,1-\varepsilon]$ for $0< \varepsilon \ll 1$. After applying successive integration by parts, and using on the left hand side the conditions \eqref{h1conds} and the asymptotic expansion in Eq.~\eqref{h1behaviour}, we get
\begin{equation}\label{beta_Eq1}
	\frac{4d}{\vartheta^3}\left(\vartheta\dot{a}_\pm\mp q_\pm\right)\left(1-\ln \varepsilon\right)-4\beta_\pm = \frac{4d}{\vartheta^3}\left[\vartheta\dot{a}_\pm\left(\ln 2-\ln \varepsilon-1\right)+\vartheta\dot{a}_\mp \pm\frac{3\dot{v}}{2d}\mp q_\pm-\tilde{I}_\pm^{\varepsilon}\right] +O(\varepsilon\ln\varepsilon),
\end{equation}
where
\begin{equation}
	\tilde{I}_{\pm}^\varepsilon = \integ{-1+\varepsilon}{1-\varepsilon}{\left[\frac{1}{2}\ln\left(\frac{1+s}{1-s}\right)\pm\frac{1}{1\mp s}\right]q}{s}.
\end{equation}
Equation \eqref{beta_Eq1} is solved for $\beta_\pm$ to find
\begin{equation}\label{beta}
	\beta_{\pm} = \frac{d}{\vartheta^2}\left[\dot{a}_{\pm}\ln\frac{\e^2}{2}-\dot{a}_\mp\right]+\frac{d}{\vartheta^3}\left[\tilde{I}_\pm^{\varepsilon} \mp\frac{3\dot{v}}{2d}\pm q_\pm\ln\varepsilon\right]+O(\varepsilon\ln\varepsilon),
\end{equation}
noting that the logarithmically diverging terms $q_\pm\ln\varepsilon$ as $\varepsilon\to 0$ are balanced with the diverging integrals $\tilde{I}^\varepsilon_\pm$. Using
\begin{equation}
	\ln\varepsilon = \ln2-\integ{-1+\varepsilon}{1-\varepsilon}{\frac{1}{1\mp s}}{s},
\end{equation}
and putting all terms under a single integral sign, a well-defined integral is formed when we take $\varepsilon \to 0$, thus obtaining the expressions for $\beta_\pm$, see Eqs.~\eqref{beta1}.

\section{Inner-region analysis}\label{InnerRegionCalc}

In the inner region, we introduce the following stretching transformation 
\begin{equation}
	h_\In = \lambda\Upsilon_\pm,\quad \xi = \pm\frac{a_\pm-x}{\lambda}\theta_\pm,
\end{equation}
which has the effect of zooming into the two contact points and allows us to retain the effects of slip~\cite{Vellingiri2011}. Hence, the governing PDE, Eq.~\eqref{GovPDE}, transforms to
\begin{equation}
	\pm\dot{a}_{\pm}\pd[\Upsilon_\pm]{\xi}+\theta_{\pm}^3\pd[]{\xi}\left[\Upsilon_\pm(\Upsilon_\pm^2+1)\pddd[\Upsilon_\pm]{\xi}\right]=\frac{q_\pm}{\theta_\pm},
\end{equation}
where we dropped $O(\lambda)$ terms, assumed that $q$ varies at lengthscales that are longer than $\lambda$, and took $|\dot{a}_\pm|,|q_\pm|\gg \lambda$. This is a generalisation of Ref.~\cite{Vellingiri2011} which treated the case of $q\equiv0$ and $\theta$ variable, and that of Ref.~\cite{OLIVER2015} which treated both $q$ and $\theta$ as constants. At $\xi=0$ we require
\begin{equation}
	\Upsilon_\pm=0,\quad \pd[\Upsilon_\pm]{\xi}=1,
\end{equation}
including $\Upsilon_\pm/\xi^2 \to 0 \text{ as } \xi \to \infty$ to ensure compatibility with the outer solution \eqref{OuterSlope}. Similar to the outer region analysis, we introduce a quasistatic expansion in the form
\begin{equation}
\Upsilon_\pm = \xi+\tilde{\Upsilon}_{\pm}+\ldots,
\end{equation}
so that $\xi \gg \tilde{\Upsilon}$, and thus obtain the following equations for $\tilde{\Upsilon}_{\pm}$
\begin{equation}
\pddd[\tilde{\Upsilon}_{\pm}]{\xi} = \frac{q_\pm \mp\dot{a}_{\pm}\theta_\pm}{\theta_{\pm}^4(\xi^2+1)},
\end{equation}
which are solved with $\tilde{\Upsilon}_\pm = \pd[\tilde{\Upsilon}_\pm]{\xi} = 0$ at $\xi = 0$ as well as $\tilde{\Upsilon}_\pm/\xi^2\to 0$ as $\xi\to\infty$. The asymptotic behavior of the slope of $\tilde{\Upsilon}_\pm$ as $\xi \to \infty$ is found to be
\begin{equation}
\pd[\tilde{\Upsilon}_{\pm}]{\xi} \sim -\frac{\left(q_\pm \mp \dot{a}_{\pm}\theta_\pm\right)}{\theta_{\pm}^4}\ln(\e\xi),
\end{equation}
from which we can write the corresponding asymptotic behavior for the inner slopes in the original variables, see Eq.~\eqref{InnerSlope}.

\section{Matching when fluxes do not vanish at the contact points}\label{MatchingFull}

As alluded to in Sec.\ \ref{Sec:MA}, the slopes \eqref{OuterSlope} and \eqref{InnerSlope} can be matched using the methods developed by Sibley and co-authors \cite{Sibley2015}. Referring the more interested reader to Ref.~\cite{Sibley2015} for more details, matching for the right/left contact points requires a different function of $\partial_xh$, denoted by $G_\pm$, respectively. These functions are given by
\begin{equation}\label{eq:G}
	G_\pm\left(\phi\right) = \pm\integ{0}{\phi}{\frac{y^3\dot{a}_\pm}{\dot{a}_\pm y\mp q_\pm}}{y}= \pm \frac{\phi^3}{3} + \frac{\phi^2q_\pm}{2\dot{a}_\pm} \pm \frac{\phi q_\pm^2}{\dot{a}_\pm^2} + \frac{q_\pm^3}{\dot{a}_\pm^3}\ln\left[\frac{\mp(\dot{a}_\pm\phi\mp q_\pm)}{q_\pm}\right].
\end{equation}
The integrand in Eq.~\eqref{eq:G} is essentially the product of $\dot{a}_\pm$ with the reciprocal of the coefficient multiplying the logarithmically diverging term in \eqref{OuterSlope} and replacing $\vartheta$ by the integration variable, $y$. Then, we find that $G_\pm(\partial_x h_\Out)$ matches directly with $G_\pm(\partial_x h_\In)$ near the right/left contact points, respectively (for the special case when $q_\pm=0$, $G_\pm(\phi)=\pm\phi^3/3$ and the usual matching-of-the cubes argument follows). After some algebra, we obtain the following transcendental equations that govern the motion of the two contact points
\begin{multline}\label{transcendental}
\pm\frac{\vartheta^3-\theta_{\pm}^3}{3}+\frac{q_\pm\left(\vartheta^2-\theta_\pm^2\right)}{2\dot{a}_\pm}\pm\frac{q_{\pm}^2\left(\vartheta-\theta_\pm\right)}{\dot{a}_\pm^2}+\frac{q_{\pm}^3}{\dot{a}_\pm^3}\ln\left(\frac{\dot{a}_\pm\vartheta\mp q_\pm}{\dot{a}_\pm\theta_\pm \mp q_\pm}\right)\\
=\dot{a}_\pm\ln\left(\frac{2d\e\theta_\pm}{\lambda}\right)-\frac{\vartheta\dot{a}_\pm}{\vartheta\dot{a}_\pm \mp q_\pm}\left(2\dot{a}_\pm-\dot{a}_\mp\mp\frac{3\dot{v}}{2d\vartheta}+\frac{I_\pm}{\vartheta}\right).
\end{multline}
Eq.~\eqref{transcendental} corresponds to a highly non-trivial contact line law. It is reminiscent of the equation obtained by Oliver \emph{et al.}\ for constant flux $q=\dot{v}/2d$ with $\theta(x) = 1$, and without the $O(1/|\ln\lambda|^2)$ terms which are included here~\cite{OLIVER2015}. Based on the arguments presented in Ref.~\cite{OLIVER2015}, we anticipate that for given values of $\vartheta\neq0$, $\theta_\pm$, $\dot{v}$, $q_\pm$ and $I_\pm$ we can solve Eq.~\eqref{transcendental} to uniquely determine the contact point velocities $\dot{a}_\pm$, even as $\dot{a}_\pm\to0$.

\bibliography{Bibliography2D}{}
\end{document}